\documentclass[fleqn,a4paper,prb,twocolumn,showpacs]{revtex4-1}
\usepackage[pdftex]{graphicx}
\usepackage[english]{babel}
\usepackage{amssymb}
\usepackage{amsmath}
\usepackage[utf8]{inputenc}
\usepackage{units}
\usepackage{color}

\newcommand{\YBCO}{YBa$_2$Cu$_3$O$_{7-\delta}$\ }
\newcommand{\ud  }{\; \mathrm{d}}

\begin{document}

\title{An analytical approach to the thermal instability of superconducting films  \\ under high current densities}
\author{Jes\'us Maza}
\email [E-mail me at: ]{jesusj.maza@usc.es}
\author{Gonzalo Ferro}
\author{Manuel Rodr\'iguez Osorio} 
\altaffiliation[Present address: ]{Laboratorio de Bajas Temperaturas, Departamento de F\'isica de la Materia Condensada, Universidad Aut\'onoma de Madrid, Spain}
\author{Jos\'e A. Veira}
\author{F\'elix Vidal}
\affiliation{Laboratorio de Bajas Temperaturas y Superconductividad, Departamento de F\'isica de la Materia Condensada, Universidad de Santiago de Compostela,  Spain}

\begin{abstract}
Using the Green's function of the 3D heat equation, we develop an
analytical account of the thermal behaviour of superconducting films
subjected to electrical currents larger than their critical current
in the absence of an applied magnetic field. Our model assumes homogeneity
of films and current density, and besides thermal coefficients employs
parameters obtained by fitting to experimental electrical field -
current density characteristics at constant bath temperature. We derive
both a tractable dynamic equation for the real temperature of the
film up to the supercritical current density $J^{*}$ (the lowest
current density inducing transition to the normal state), and a thermal
stability criterion that allows prediction of $J^{*}$. For two typical
YBCO films, $J^{*}$ predictions agree with observations to within
5\%. These findings strongly support the hypothesis that a current-induced
thermal instability is generally the origin of the breakdown of superconductivity
under high electrical current densities, at least at temperatures
not too far from $T_{c}$. 
\end{abstract}

\maketitle

\section{INTRODUCTION}

In experiments in which superconducting films are placed in an environment
({}``bath'') that is thermostatted at a temperature $T_{b}$ below
their critical temperature $T_{c}$, and are then subjected to a gradually
increasing current density $J$, the electric field becomes measurable
at the critical current density $J_{c}(T_{b})$. It then increases
ever more rapidly with $J$ until, at the {}``supercritical'' or
{}``quench'' current density $J^{*}(T_{b})$, it jumps to values
corresponding to the nonsuperconducting state, the discontinuity at
$J^{*}$ becoming increasingly abrupt as the bath temperature is lowered.
This phenomenon, which is of interest not only for the theory of electrical
transport in superconductors but also in relation to some of their
most important applications, is still poorly understood. The main
mechanisms proposed so far may be crudely classified in two classes.
One, comprising what may be termed current-driven mechanisms, basically
invokes electrodynamic effects dependent on the microstructure of
the sample.\cite{Doettinger94,Kunchur02,Babic04,Bezuglyi10,Bernstein07}
The other invokes heat-driven mechanisms that are essentially artifactual,
postulating that a small increase in temperature due to the finite
duration of measurements triggers a thermal runaway.\cite{Lindmayer93,Kiss99,Lehner02,Maza08}
(Further discussion of both approaches is available.\cite{Maza08}) 

A weakness of studies exploring the heat-driven account has hitherto
been their reliance on results that were obtained by numerical methods,
 the limited scope of which somewhat obscures their theoretical
interpretation. In this paper we address this weakness by developing
an analytical theory of the thermal stability of high-$T_{c}$ films
that explains previous experimental and simulational results.\cite{Maza08}
The theory presented is a full 3D model that provides a dynamic equation
for the temperature of the film as a function of time, together with
a thermal stability criterion with clear-cut predictions for the dependence
of $J^{*}$ on bath temperature and film geometry. Its parameters
are those of a homogeneous film material; no appeal is made to hard-to-quantify
microstructural defects, which in some other models play the role
of free parameters that facilitate good fit to experimental results. 

We know of no previous studies that significantly overlap with this
work. In particular, the monumental review by Gurevich and Mints\cite{Gurevich87}
deals only briefly with the thermal stability of homogeneous superconductors,
and then only for thin wires and at the hard superconductivity limit
($J_{c}\gg J-J_{c})$, conditions that are far removed from those
considered here.

\section{BACKGROUND THEORY: HEATING AN INFINITE MEDIUM}

Consider a point source embedded at $\vec{r}_0=(\xi,\eta,\zeta)$
in an infinite homogeneous medium that at time $t=0$ has zero temperature.
The evolution of the temperature field $T$ following delivery of
a heat pulse at time $t=t_{0}$ is governed by the heat equation 

\begin{equation}
\nabla^{2}T-\frac{1}{D}\frac{\partial T}{\partial t}=-4\pi\delta(\vec{r}-\vec{r}_0)\delta(t-t_{0})\label{eq:greeneq}\end{equation}
 where $D$ is the thermal diffusivity of the medium. The solution
is\cite{Morse53}

\begin{equation}
T_{G}(\vec{r},t\vert\vec{r}_0,t_{0})=\left\{ \begin{array}{ll}
0 & t<t_{0}\\
\frac{4\pi D^{2}}{\{4\pi D(t-t_{0})\}^{3/2}}\,\exp\left(-\frac{(\vec{r}-\vec{r}_0)^{2}}{4D(t-t_{0})}\right)\,\,\,\,\,\,\,\,\, & t>t_{0}\end{array}\right.\label{eq:greensol}\end{equation}

For a general heating rate density $\dot{Q}\left(\vec{r},t\right)$
, substitution of Fourier's law in the heat balance equation

\begin{equation}
c\frac{\partial T\left(\vec{r},t\right)}{\partial t}+\vec{\nabla}\cdot\vec{q}\left(\vec{r},t\right)=\dot{Q}\left(\vec{r},t\right)\label{eq:heat}\end{equation}
(where $\vec{q}$ is the heat flux and $c$ the specific heat at constant
pressure per unit volume) affords 

\begin{equation}
\nabla^{2}T(\vec{r},t)-\frac{1}{D}\frac{\partial T}{\partial t}(\vec{r},t)=-\frac{1}{\kappa}\dot{Q}(\vec{r},t)\label{eq:temp}\end{equation}
where the thermal conductivity $\kappa=cD$. The Green's function
for solution of Eq.~(\ref{eq:temp}) is $T_{G}$ [Eq.~(\ref{eq:greensol})]: 

\begin{equation}
T(\vec{r},t)=\int_{\mathcal{V}}\int_{0}^{t}\frac{\dot{Q}(\vec{r}_0,t_{0})}{4\pi\kappa}T_{G}(\vec{r},t\vert\vec{r}_0,t_{0})\ud t_{0}\ud^{3}\vec{r}_0+T_{b}\label{eq:tempsum}\end{equation}
where $T_{b}$ is the initial temperature of the medium and $\mathcal{V}$
is a region containing all $\left(\vec{r}_0,t_{0}\right)$ for which
$\dot{Q}\left(\vec{r}_0,t_{0}\right)$ is nonzero.

\section{THE MODEL}

\subsection{Constructing the model}

\label{sect:feat} 
\begin{figure}[ht]
 \includegraphics[width=0.35\textwidth]{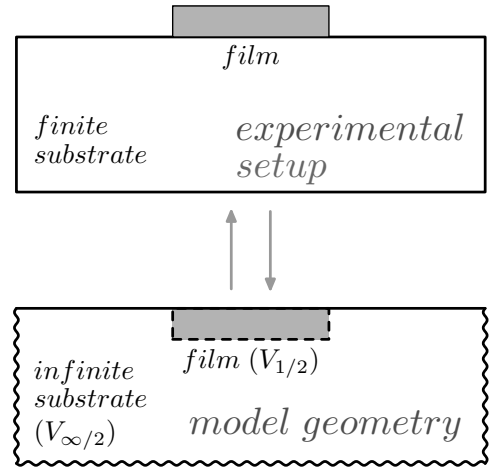}
\caption{\label{fig:expmod}Comparison of the model geometry with the schematic
experimental setup. In the model, film and substrate are of the same
material so as to be able to use Eq.~(\ref{eq:tempsum}). }
\end{figure}

\begin{figure}[ht]
 \includegraphics[width=0.4\textwidth]{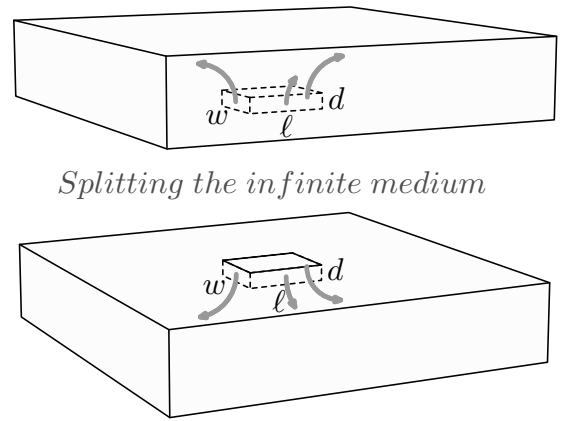}
 \caption{\label{fig:splitting}The model and its mirror image. Heat flow (arrows) does not cross the interface between model and image.}
\end{figure}

We model the film as an homogeneous parallelepiped
${\mathcal V}_{\nicefrac{1}{2}}$ of dimensions $\ell\times w\times d$
embedded in the face of a semi-infinite region $\mathcal{V}_{\nicefrac{\infty}{2}}$
of the same material with no discontinuity between the film and this
substrate (Fig.\,\ref{fig:expmod}); and we assume that heat generated
in the film flows only into the substrate, not into the overlying
refrigerant, so that heat flux through the face of $\mathcal{V}_{\nicefrac{\infty}{2}}$
and the free face of $\mathcal{V}_{\nicefrac{1}{2}}$ is identically
zero. This allows application of the method of images\cite{Barton95}:
if $\mathcal{\bar{V}}_{\nicefrac{1}{2}}$ and $\mathcal{\bar{V}}_{\nicefrac{\infty}{2}}$
are the mirror images of $\mathcal{V}_{\nicefrac{1}{2}}$ and $\mathcal{V}_{\nicefrac{\infty}{2}}$
in the plane of their free faces (Fig.\,\ref{fig:splitting}), we
need only perform calculations for $\mathcal{V}$, the union of $\mathcal{V}_{\nicefrac{1}{2}}$
and $\mathcal{\bar{V}}_{\nicefrac{1}{2}}$; and since $\mathcal{V}$
is a region of a homogeneous infinite medium, and contains all the
sources heating this medium, we can use Eq.~(\ref{eq:tempsum}) to
solve our problem.

Various features of the proposed model invite justification. In the
first place, the assumption that heat generated in the film flows
only into the substrate is an acceptable approximation if the overlying
refrigerant is gaseous, and also if liquid nitrogen is used, since
the thermal conductivity of the solid substrate can easily be two
orders of magnitude greater than that of liquid nitrogen.\cite{Bejan03}
Secondly, the distortion introduced by the film being embedded in
the substrate, rather than lying upon it, must be negligible, since
the thickness of the film is far smaller than its width or length.
Thirdly, the error introduced by treating the substrate as semi-infinite
must be negligible, because substrate and film dimensions typically
differ by about three orders of magnitude. Fourthly, treating substrate
and film as being different regions of the same infinite piece of
homogeneous material means that the thermal impedance between the
two is zero, whereas the accepted value of the actual film-substrate
exchange coefficient is $h=10^{3}$ W/K cm;\cite{Nahum91,Marshall93,Li99}
that this approximation is acceptable is shown by previous work in
which it made little difference to the results of numerical calculations.\cite{Maza08}
Finally, treating substrate and film as being made of the same material
also means that they have the same diffusivity; this will be handled
by choosing a diffusivity coefficient in between that of the real
film and the real substrate.

Our next simplification is to assume that the heating rate in the
film does not depend on position, but only on time: $\dot{Q}\equiv\dot{Q}\left(t\right)$.
Since the heating of the films we are considering will be caused by
electrical current, this assumption is equivalent to assuming that
the current density is uniform throughout the film. Explicit support
for this comes from the work of Herrmann et al.,\cite{Herrmann98,Herrmann99}
who found that in high-T$_{c}$ tapes current in excess of a certain
characteristic cutoff (the value of which was slightly below the critical
current) was indeed distributed homogeneously across the entire superconductor
cross section. Moreover, direct measurements on a slab of Bi-based
crystal, using a microarray of Hall probes, show that although current
is restricted to the lateral regions of the slab at relatively low
temperature ($T\lesssim50$\,K), nonuniformity becomes negligible
above about 80~K, a temperature relatively near $T_{c}$.\cite{Fuchs98}
Again, a recent study\cite{Bobyl02} of {Y}{B}a$_{2}${C}u$_{3}${O}$_{7-\delta}$
strips using magneto-optical imaging found that even at so low a temperature
as 20\,K the nonuniformity of $J$ is only of the order of 20\% when
the transport current is 90\% of its critical value. Finally, a uniform
distribution of current throughout the cross section is the simplest
explanation of the observation that critical current is independent
of bridge width.\cite{Kunchur00,Ruibal07,Hahn95,Dinner07}

Given uniform heating and the homogeneity of film and substrate (and
since no measurements of local temperatures have yet been made on
standard high-$T_{c}$ bridges), we are interested only in the volume-averaged
temperature of the film. From Eq.~(\ref{eq:tempsum}), 

\[
T(t)=T_{b}+\frac{1}{V}\int_{\mathcal{V}}\int_{\mathcal{V}}\int_{0}^{t}\frac{\dot{Q}(t_{0})}{4\pi\kappa}T_{G}(\vec{r},t\vert\vec{r}_0,t_{0})\ud t\ud^{3}\vec{r}_0\ud^{3}\vec{r}\]

\[
=T_{b}+\int_{\mathcal{V}}\int_{0}^{t}\frac{\dot{Q}(t_{0})}{4\pi\kappa}\times\Bigl[\frac{\sqrt{\pi}\lambda}{4d}\left(\mathrm{erf}\,\frac{d-\eta}{\lambda}+\mathrm{erf}\,\frac{d+\eta}{\lambda}\right)\times\]

\[
\,\,\,\,\,\,\,\,\,\,\,\,\,\,\,\,\,\times\frac{\sqrt{\pi}\lambda}{2w}\left(\mathrm{erf}\,\frac{w/2-\xi}{\lambda}+\mathrm{erf}\,\frac{w/2+\xi}{\lambda}\right)\times\]

\[
\,\,\,\,\,\,\,\,\,\,\,\,\,\,\,\,\,\times\frac{\sqrt{\pi}\lambda}{2\ell}\left(\mathrm{erf}\,\frac{\ell/2-\zeta}{\lambda}+\mathrm{erf}\,\frac{\ell/2+\zeta}{\lambda}\right)\Bigr]\ud t_{0}\,\ud\vec{r}_0\]

\begin{equation}
=T_{b}+\int_{0}^{t}\frac{\dot{Q}(t_{0})}{c}\, M(t-t_{0})\ud t_{0}\label{eq:fulldyneq}\end{equation}
where $V$ is the volume of $\mathcal{V}$, $\lambda=2\sqrt{D(t-t_{0})}$
is the diffusion length, erf is the error function, and 

\begin{eqnarray}
 &  & M(t-t_{0})=[\mathrm{erf}\,\frac{w}{\lambda}+\frac{\lambda}{\sqrt{\pi}w}(e^{-\frac{w^{2}}{\lambda^{2}}}-1)]\times\nonumber \\
 &  & \,\,\,\,\,\,\,\,\,\,\,\,\,\,\times[\mathrm{erf}\,\frac{\ell}{\lambda}+\frac{\lambda}{\sqrt{\pi}\ell}(e^{-\frac{\ell^{2}}{\lambda^{2}}}-1)]\times\nonumber \\
 &  & \,\,\,\,\,\,\,\,\,\,\,\,\,\,\times[\mathrm{erf}\,\frac{2d}{\lambda}+\frac{\lambda}{\sqrt{\pi}2d}(e^{-\frac{4d^{2}}{\lambda^{2}}}-1)]\label{eq:M}\end{eqnarray}

\begin{figure}[ht]
 \includegraphics[width=0.4\textwidth]{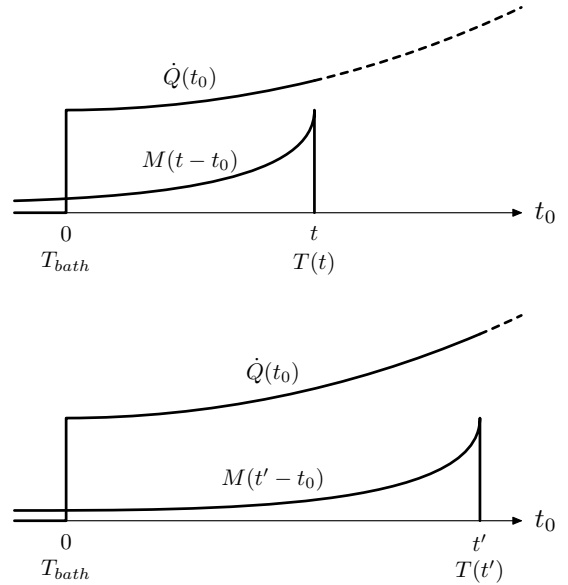}
\caption{\label{fig:conv}Schematic diagrams of the heating rate $\dot{Q(t_{0})}$
and the transfer function $M(t-t_{0})$ at two times, $t$ and $t'>t$. }
\end{figure}

\begin{figure}[hb]
 \includegraphics[width=0.4\textwidth]{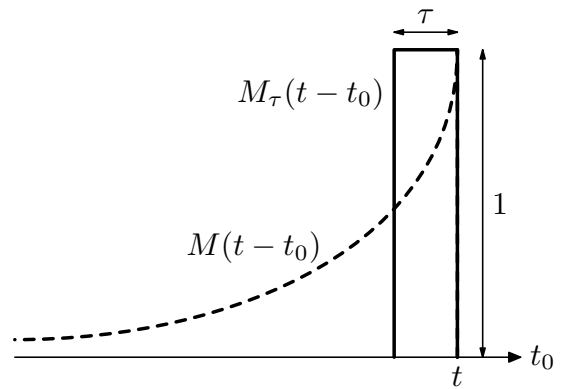}
\caption{\label{fig:Mtc}Comparison of the transfer function $M(t-t_{0})$
with its approximation, $M_{\tau}(t-t_{0})$. }

\end{figure}

Note that although the precise shape of the transfer function\cite{Rabiner75} \cite{Born75}
$M$ depends on the dimensions and diffusivity of the film, its general
shape is as shown (reversed) in Fig.\,\ref{fig:conv}. This means
that in the final expression in Eq.~(\ref{eq:fulldyneq}), the convolution\cite{Oran88}
of $\dot{Q}(t_{0})$ with $M(t-t_{0})$ gives considerably more weight
to the immediate past and the present moment ($\underset{t_{0}\rightarrow t}{\lim}M(t-t_{0})=1$)
than to the remote past ($\underset{t_{0}\rightarrow-\infty}{\lim}M(t-t_{0})=0$)
- as in fact was only to be expected. We exploit this behaviour by
making a further approximation consisting in the replacement of $M$
as defined in Eq.~(\ref{eq:M}) by $M_{\tau}$, a rectangular function
of unit height that is non-zero on the interval $[0,\tau]$, where
$\tau$ is the area under $M$ (Fig.\,\ref{fig:Mtc}):

\begin{equation}
T(t)=T_{b}+\int_{0}^{t}\frac{\dot{Q}(t_{0})}{c}\, M_{\tau}(t-t_{0})\ud t_{0}=T_{b}+\int_{t-\tau}^{t}\frac{\dot{Q}(t_{0})}{c}\ud t_{0}\label{eq:MtauT}\end{equation}

We call $\tau$ the characteristic time of the film. Furthermore,
since $M_{\tau}$ clearly overweights the proximal past, we slightly
correct the integral by taking $\dot{Q}(t_{0})$ to have the value
$\dot{Q}(t-\tau)$ for all $t_{0}$ in $[t-\tau,t]$, i.e. we replace
$M_{\tau}(t-t_{0})$ with $M_{\delta}=\tau\delta(t-\tau-t_{0})$.
Thus, finally (insofar as the dependence of temperature on time via
the heating rate), $T(t)=\tau\dot{Q}(t-\tau)/c+T_{b}$, or

\begin{equation}
T(t+\tau)=\tau\frac{\dot{Q}[J,E(J,T(t))]}{c}+T_{b}\label{eq:tcdyneq}\end{equation}
where the origin of the time-dependence of $\dot{Q}$ is explicitly
displayed as the temperature- (and hence time-) dependence of $E(J)$,
the intrinsic current-voltage characteristic (CVC) of the film. 

\begin{figure}[ht]
 \includegraphics[width=0.45\textwidth]{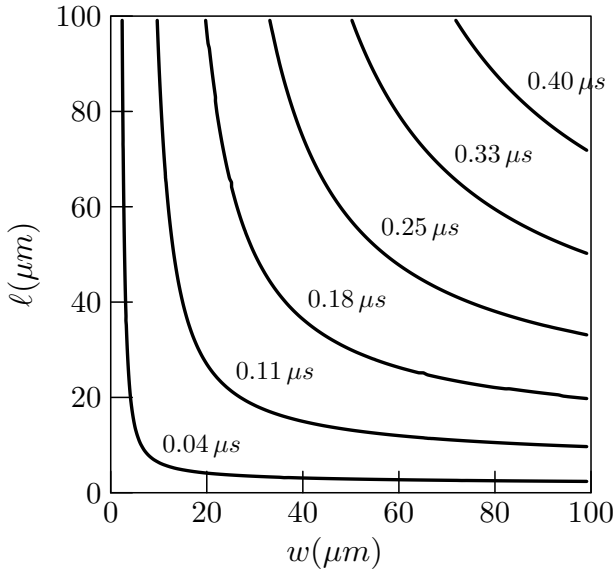}
\caption{\label{fig:tclines}Isolines of the characteristic time $\tau$ of
a typical film 0.15 $\mathrm{\mu m}$ thick as a function of film
width and length. }
\end{figure}

Because of the structure of $M$ as a product of factors that each
depend on only one spatial dimension (as well as on time, through
$\lambda$), $\tau$ tends to zero whenever $w$,
$l$ or $d$ do and the others are held fixed, and tends to the product
of the other two factors whenever $w$, $l$ or $d$ tend to infinity;
the larger $w$, $l$ and $d$, the larger is $\tau$, i.e. the greater
the thermal inertia of the film, though $\tau$ always lies in {[}0,1{]}.
That for a fixed thickness and width/length ratio the thermal behaviour
of the film does depend on width, but increasingly less as width increases,
has recently been confirmed experimentally.\cite{Ruibal07} The dependence
of $\tau=\int_{0}^{\infty}M(t')\ud t'$ on film width and length for
a typical thickness (0.15 $\mathrm{\mu m}$) is shown as a contour
map in Fig.\,\ref{fig:tclines}. 

Note that if with a fixed current density $J$, $T$ tends to a stable
value $T_{\infty}$, then Eq.~(\ref{eq:tcdyneq}) is asymptotically
equivalent to Eq.~(\ref{eq:MtauT}), because $\dot{Q}$ must then
also be stable and can therefore be taken out from under the integral
in Eq.~(\ref{eq:MtauT}).

Since our heat source is in this study the Joule effect of a constant
current, the form of $\dot{Q}[J,E(J,T)]$ in Eq.~(\ref{eq:tcdyneq})
is 

\begin{equation}
\dot{Q}[J,E(J,T)]=JE(J,T)\label{eq:elecpow}\end{equation}
For $E(J,T)$, which is not known (the experimental CVCs at constant
bath temperature showing the very distortion that is investigated
in this study), we shall use two empirical functions with different
forms in order to show that our results do not depend critically on
this aspect of the model. The first, variants of which have been used
in previous work by ourselves\cite{Vinha03,Maza08} and others,\cite{Solovjov94,Fisher85,Prester98}
is 

\begin{equation}
E_{n}(J,T)=E_{0}(T)\,\big[\frac{J}{J_{0}(T)}-1\big]^{n}\label{eq:powlawn}\end{equation}
 where $E_{0}(T)=E_{01}(1-T/T_{c})^{m}$ and $J_{0}(T)=J_{01}(1-T/T_{c})^{m}$.
The second is

\begin{equation}
E_{s}(J,T)=\rho_{n}(T)\,\big[J^{\frac{1}{s}}-J_{0}(T)^{\frac{1}{s}}\big]^{s}\label{eq:powlaws}\end{equation}
where $J_{0}(T)$ is as in Eq.~(\ref{eq:powlawn}), $s=s_{0}+s_{1}(1-T/T_{c})$,
and $\rho_{n}(T)$ is obtained by extrapolation from data for the
temperature dependence of the resistivity of the normal (non-superconducting)
film. Both have four adjustable parameters ($E_{01},\, J_{01},\, m\,\mathrm{and}\, n$
in Eq.~(\ref{eq:powlawn}), $J_{01},\, m,\, s_{0}\,\mathrm{and}\, s_{1}$
in Eq.~(\ref{eq:powlaws}) and, as in previous work, both are to
be fitted to experimental $(J,E)$ data for the region of small $E$,
where little heat is generated and the experimental CVC can accordingly
be expected to lie close to the intrinsic CVC. We shall call the isotherms
corresponding to $E_{n}$ and $E_{s}$, \emph{n}-isotherms and \emph{s}-isotherms,
respectively. It may be noted that whereas $E_{s}$ tends to the natural
limit $\rho_{n}J$ when $J\gg J_{0}$ or $T\longrightarrow T_{c}$,
this is not so for $E_{n}$.

\subsection{Parameterizing and testing the model}

\begin{figure}[hb]
 \includegraphics[width=0.45\textwidth]{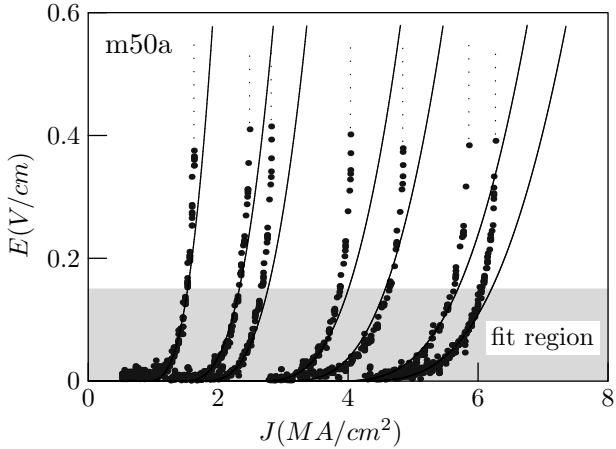}
\caption{\label{fig:isother}Large dots show experimental current-voltage characteristics
(CVCs) of a 50$\mu$m-wide \YBCO film at bath temperatures $T_{b}$
of (from right to left) 72.1, 73.5, 76.5, 78.0, 81.2, 82.3 and 84.2
K. Vertical dotted lines indicate voltage jumps at $J^{*}(T_{b})$.
Continuous curves show \emph{s-}isotherms [Eq.~(\ref{eq:powlaws})]
fitted to the data with \emph{E} values in the shaded region. }
\end{figure}

The films whose thermal behaviour we used to test the model and its
predictions were two \YBCO bridges on SrTiO$_{\text{3}}$ substrates.
One (sample \emph{mA}) was a $50\times10\times0.12\,\mu\textrm{m}$
film with $T_{c}$=89.8 K, $\rho_{n}$(100 K)\,=\,117\,$\mu\Omega$cm
and $\rho_{n}$(300 K)\,=\,374\,$\mu\Omega$cm; other features
have been published elsewhere.\cite{Vinha03} The other (sample \emph{m50a})
was longer and wider ($500\times50\times0.12\,\mu\mathrm{m}$), with
$T_{c}$=87.1 K, $\rho_{n}$(100 K)\,=\,190\,$\mu\Omega$cm, and
$\rho_{n}$(300 K)\,=\,490\,$\mu\Omega$cm.\cite{Ruibal07} By
way of illustration, Fig.~\ref{fig:isother} shows the \emph{s}-isotherms\emph{
}fitted by least squares to experimental $(J,E)$ data for sample
\emph{m50a.} 

The diffusivity $D$ of the superconducting bridge material is $0.05\,\mathrm{cm^{2}/s}$, and that of the SrTiO$_{\text{3}}$ substrate $0.18\,\mathrm{cm^{2}/s}$.\cite{Touloukian70}
The intermediate value to be used in the model (see the second paragraph
of the previous section) was informally optimized to afford adequate
fit between the experimental values of $J^{*}(T_{b})$ for sample
\emph{m50A} and predicted values that were obtained as follows. 

Although $J^{*}(T_{b})$ is defined as the current density at which
a voltage jump occurs, Eqs.~(\ref{eq:tcdyneq})-(\ref{eq:powlaws})
show that this voltage jump will be accompanied by a temperature jump.
Thus $J^{*}(T_{b})$ may be predicted by using these equations to
simulate an experimental plot of temperature against current until
a discontinuity occurs. Starting at a bath temperature $T_{b}$ at
time 0, the temperature $T(\tau)$ attained after applying a current
density $J_{c}+\delta J$ for time $\tau\ll\delta t$ is given by
Eqs.~(\ref{eq:tcdyneq}), (\ref{eq:elecpow}) and either (\ref{eq:powlawn})
or (\ref{eq:powlaws}) (where $\delta J$ is the current density step
used in the experiments, typically 0.05\,$\mathrm{MA}/\mathrm{cm^{2}}$,
and $\delta t$ the step length, typically 1\,ms); $T(2\tau)$ is
similarly obtained using $T(\tau)$ as starting temperature; and so
on until the total time elapsed is $\delta t$, whereupon the current
density is increased by $\delta J$, etc. The current density at which
a sudden temperature jump is observed is identified as $J^{*}(T_{b})$.

\begin{figure}[hb]
 \includegraphics[width=0.45\textwidth]{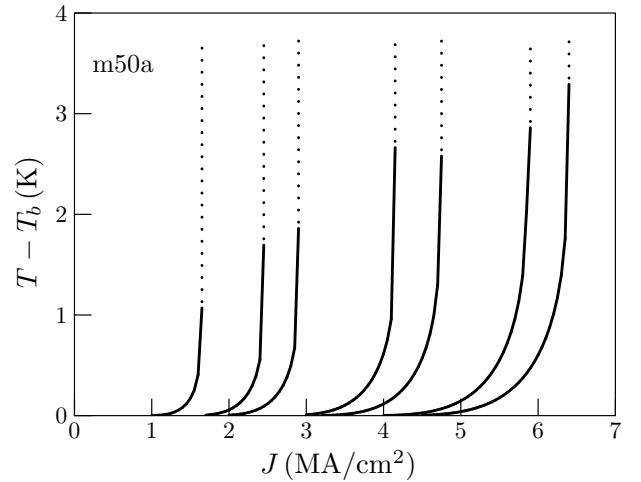}
\caption{\label{fig:DTJ_tc}Excess temperature of sample \emph{m50A}, relative
to the bath temperature $T_{b}$, as calculated using Eq.~(\ref{eq:tcdyneq})
and $s$-isotherms [Eq.~(\ref{eq:powlaws})] in simulations in which
the current density $J$ was increased by 0.05\,$\mathrm{MA}/\mathrm{cm^{2}}$ at
1\,ms intervals. From right to left, $T_{b}=$ 72.1, 73.5, 76.5,
78.0, 81.2, 82.3 and 84.2 K. A temperature jump occurs 1-3 K
above the bath temperature, the excess temperature decreasing with
increasing $T_{b}$. }
\end{figure}

Fig.\,\ref{fig:DTJ_tc} shows plots of $(T-T_{b})$ against $J$
obtained in this way using \emph{s}-isotherms, and Table\,\ref{tab:J*m50a}
confirms good fit between the predicted values of $J^{*}$ (and those
predicted similarly using \emph{n}-isotherms) and the experimental
values. The achievement of such good fit with such a crude model is
possibly attributable partly to the fact that for each $T_{b}$ the
value of $J^{*}(T_{b})$ lies only about 20\% above the largest $J$
in the set of $(J,E)$ data to which Eqs.~(\ref{eq:powlawn}) and
(\ref{eq:powlaws}) were fitted, i.e. extrapolation was quite limited;
partly to the temperature and voltage jump occurring only 1-3\,K
above $T_{b}$ (see Fig.\,\ref{fig:DTJ_tc}), i.e. far from $T_{c}$;
and, given these circumstances, to the above-noted equivalence of
Eqs.~(\ref{eq:tcdyneq}) and (\ref{eq:MtauT}) for $\tau\ll\delta t$
and $J_{c}<J<J^{*}$.

The value of $D$ affording the above results was 0.12\,$\mathrm{cm^{2}}/\mathrm{s}$;
that this is closer to the diffusivity of the subtrate than to that
of the superconductor seems reasonable, since the thermal diffusion
length of the substrate for a time of 1\,ms is rather more than 250\,$\mu\mathrm{m}$,
which is much larger than the film. The same value of $D$ also performed
well for sample \emph{mA}. The corresponding values of $\tau$ are
0.15\,$\mu\mathrm{s}$ for film \emph{mA} and 0.55~$\mu\mathrm{s}$
for \emph{m50A}; as required by the above algorithm, these values
are both much shorter than the experimental step length, 1\,ms. Note
that according to Eq.~(\ref{eq:tcdyneq}) the characteristic time
$\tau$ is the time taken by the film to return to the bath temperature
$T_{b}$ when heating is stopped. Accordingly, the use of current
pulses lasting just a few tenths of a microsecond\cite{Harrabi09}
and separated by intervals of similar length should allow the measurement
of current-voltage curves that are nearly free from artifactual thermal
effects.

\begin{table}[th]
\caption{\label{tab:J*m50a}Experimentally observed supercritical currents
of sample \emph{m50a} at seven bath temperatures, together with the
values obtained in simulations of experimental runs using Eq.~(\ref{eq:tcdyneq})
and either $n$-isotherms [$J_{n}^{*}$; Eq.~(\ref{eq:powlawn})
or $s$-isotherms $J_{s}^{*}$; Eq.~(\ref{eq:powlaws})].}

\begin{ruledtabular}
\begin{tabular}{|l l l l|}

$T_b(K)$ & $J_{exp}^* \, (\frac{MA}{cm^2})$ &  $J_n^* \, (\frac{MA}{cm^2})$ & $J_s^* \, (\frac{MA}{cm^2})$ \\
[1.2ex]
\hline
72.1& 6.27 & 6.50 & 6.50\\
73.5 & 5.86 & 5.95 & 5.95 \\
76.5 & 4.84 & 4.80 & 4.80 \\
78.0 & 4.04 & 4.20 & 4.25 \\
81.2 & 2.82 & 2.90 & 2.95 \\
82.3 & 2.49 & 2.40 & 2.50 \\
84.2 & 1.61 & 1.55 & 1.65 \\

\end{tabular}
\end{ruledtabular}
\end{table}

\section{Stability }

\begin{figure}[ht]
 \includegraphics[width=0.45\textwidth]{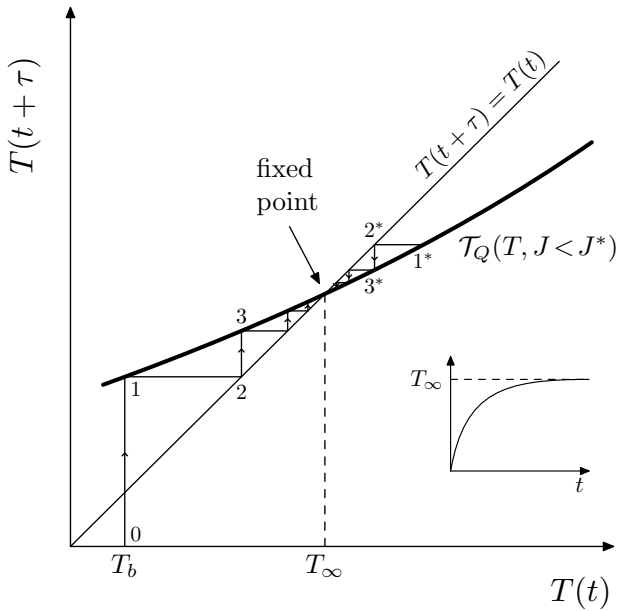}
\caption{\label{fig:stableJ}Staircase diagram of thermal dynamics under an
electrical current density $J<J^{*}$. The inset shows, as a function
of time, the approach to the stable limiting temperature $T_{\infty}$
from below.}

\end{figure}

\begin{figure}[ht]
 \includegraphics[width=0.45\textwidth]{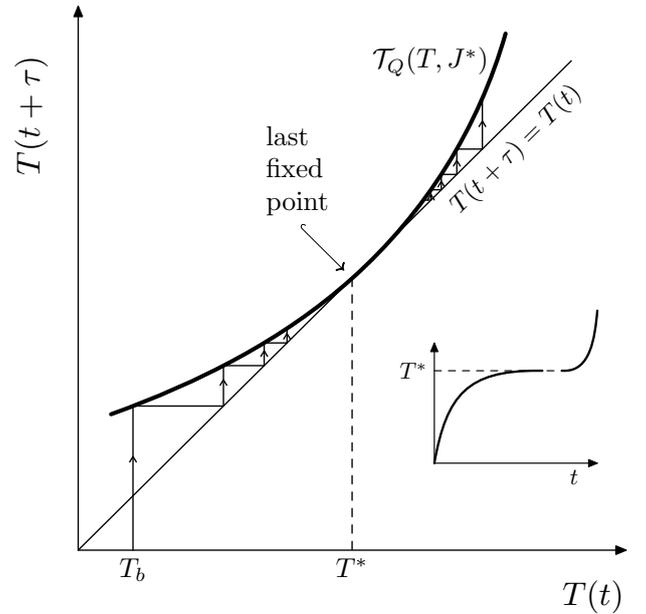}
\caption{\label{fig:jump}Staircase diagram of thermal dynamics under an electrical
current density $J=J^{*}$. Starting at low values, temperature increases
towards the asymptotically limiting value $T^{*}$, but this limit
is unstable: starting from higher values, temperature increases indefinitely. }

\end{figure}

\begin{figure}[ht]
 \includegraphics[width=0.45\textwidth]{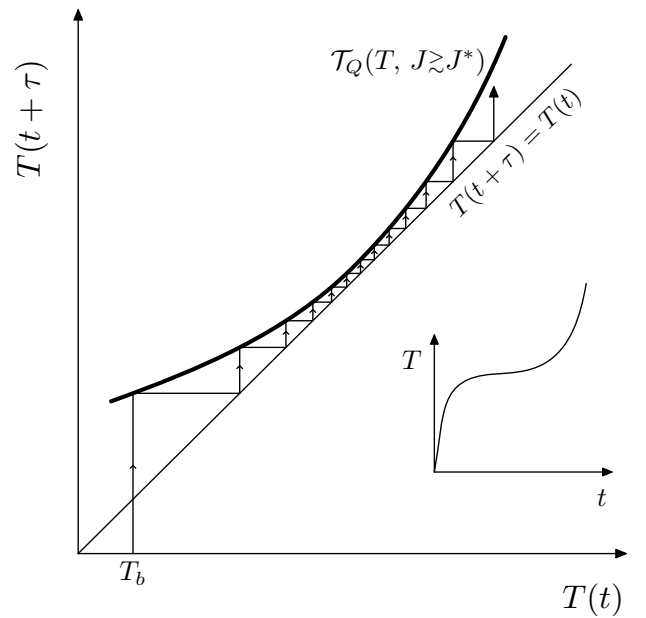}
\caption{\label{fig:unstable}Staircase diagram of thermal dynamics under an
electrical current density $J>J^{*}$. The rate of increase of temperature
slows down, but finally increases without limit.}

\end{figure}

Eq.~(\ref{eq:tcdyneq}) is a\emph{ }nonlinear autonomous difference
equation of first order.\cite{Elaydi99} To examine the stability
of $T$ at constant $T_{b}$ and under a fixed current density $J$
we rewrite this equation in the form 

\begin{equation}
T(t+\tau)=\mathcal{T}_{Q}[T(t);J]\label{eq:tcdyneq2}\end{equation}
where $\mathcal{T}_{Q}[T(t);J]\equiv\tau\,\frac{{Q}[T(t),J]}{c}+T_{b}$,
and we note that under the given conditions $\mathcal{T}_{Q}(T;J)$
is convex and monotonically increasing, and that $\mathcal{T}_{Q}(T_{b};J)>T_{b}$
(since the current must heat the film). The condition for attainment
of a stable temperature $T_{\infty}$ is that the graph of $\mathcal{T}_{Q}(T;J)$
intersect the line $T(t+\tau)=T(t)$, as may be seen by examining
the staircase diagram\cite{Elaydi99} shown in Fig.\,\ref{fig:stableJ},
in which vertical arrowed lines indicate real changes in temperature
during a time increment $\tau$, while horizontal lines translate
the final temperature of one $\tau$-interval into the starting temperature
of the next. Not only does the path $0\rightarrow1\rightarrow2\rightarrow3\rightarrow\cdots$
lead from its starting point ($T_{b}$) to the limiting temperature
$T_{\infty}$ (a fixed point of $\mathcal{T}_{Q}(T;J)$), but
so does the path $1^{*}\rightarrow2^{*}\rightarrow3^{*}\rightarrow\cdots$,
i.e. any fluctuation to a temperature higher than $T_{\infty}$ will
be recovered from. By contrast, if $\mathcal{T}_{Q}(T;J)$ meets
the line $T(t+\tau)=T(t)$ tangentially the limiting temperature $T^{*}$
is not stable (Fig.\,\ref{fig:jump}), and if $\mathcal{T}_{Q}(T;J)$
lies wholly above $T(t+\tau)=T(t)$ there is no limiting temperature
(Fig.\,\ref{fig:unstable}). 

It may be enlightening to compare the above situations with that of
a normal conductor, for which $E(T,J)=\rho(T)J$, where the electrical
resistivity $\rho(T)=\rho_{0}+\rho_{1}T$. $\mathcal{T}_{Q}(T;J)$
is in this case a linear function of temperature, and whether it intersects
the line $T(t+\tau)=T(t)$ (i.e. whether a stable limiting temperature
is attained) therefore depends on the slope of this linear function.
For copper, for example, $\partial\mathcal{T}_{Q}/\partial T\lesssim10^{-3}$
for $J=1$ MA/cm$^{2}$ and $\tau=0.14$ $\mu$s,\cite{Buch99}  so
the absence of thermal runaway is ensured. 

Since $\mathcal{T}_{Q}(T,J)$ increases with $J$ [by Eq.~(\ref{eq:elecpow})],
increasing the current density with a given bath temperature results
in the location of the curve $\mathcal{T}_{Q}(T;J)$ progressing
from that shown in Fig.\,\ref{fig:stableJ} to that shown in Fig.\,\ref{fig:unstable};
and examination of the definition of $\mathcal{T}_{Q}(T;J)$
shows that the greater the thermal inertia of the superconducting
film (i.e. the greater $\tau$), the smaller the current at which
instability sets in. The supercritical current density $J^{*}$ is
the current density corresponding to Fig.\,\ref{fig:jump}, a situation
that can be characterized by the conditions 

\begin{eqnarray}
 &  & T^{*}=\mathcal{T}_{Q}(T^{*};J^{*})\nonumber \\
 &  & \frac{\partial\mathcal{T}_{Q}}{\partial T}(T^{*};J)=1\label{eq:jump}\end{eqnarray}
Alternatively, $J^{*}$ can be characterized as the smallest current
density such that: \begin{equation}
\mathcal{T}_{Q}(T;J)\ge T,\quad\forall\; T>T_{b}\label{eq:jump2}\end{equation}
(and the supercritical temperature $T^{*}$ as the temperature at
which equality holds), or as the largest current density for which
the equation \begin{equation}
\mathcal{T}_{Q}(T;J)\,=\, T\label{eq:jump3}\end{equation}
has a solution. 

To ensure the identification of $J^{*}$ for a given $T_{b}$, we first
find the solution $T_{\infty}$ of Eq.~(\ref{eq:jump3}) for a value
of $J$ just slightly greater than $J_{c}$, then for a slightly larger
$J$, and so on, until the largest $J$ for which there is a solution
is identified. This procedure is not a simulation analogous to those
of Section\,3, for example, because the criterion for increasing
$J$ is not the time elapsed but the satisfaction of a criterion of
convergence that cannot be verified experimentally (at least at present).
In experimental practice and simulations, $J$ is not kept fixed indefinitely;
if $J$ is increased too fast, $J^{*}$ will be missed, though the
thermal runaway will of course occur (alternatively, if $\mathcal{T}_{Q}(T;J)$
approaches $T(t+\tau)=T(t)$ quite closely in the situation of Fig.\,\ref{fig:unstable},
the slowing down of the increase in temperature may be erroneously
taken to indicate approach to a non-existent $T_{\infty}$). 

\begin{figure}[ht]
 \includegraphics[width=0.48\textwidth]{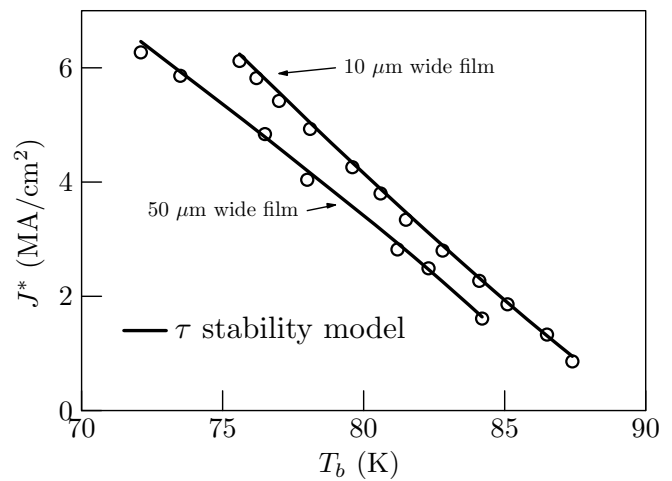}
\caption{\label{fig:jstar}Predictions of $J^{*}$ for samples \emph{mA} and
\emph{m50A} obtained using Eq.~(\ref{eq:tcdyneq}), $s$-isotherms
[Eq.~(\ref{eq:powlaws})] and the convergence illustrated in Fig.\,\ref{fig:stableJ}
for $J$ values that were successively increased until convergence
failed. The experimental points correspond to the abrupt voltage jumps
observed in experimental CVCs (for sample \emph{m50A}, Fig.~\ref{fig:isother}).}

\end{figure}

For the films considered in Section\,3, Fig.\,\ref{fig:jstar} shows
the good fit between the $J^{*}(T_{b})$ functions obtained as above
and the experimental data. Experimental error may reasonably be regarded
as negligible (at least at this representation scale), because the
current at which the voltage jump takes place is quite well defined,
the current-voltage curve being locally nearly vertical. That $J^{*}(T_{b})$
lies at higher values for sample \emph{mA} than for \emph{m50A} is
expected because of its smaller characteristic time and correspondingly
lower $\mathcal{T}_{Q}(T;J)$. 

\begin{figure}[ht]
 \includegraphics[width=0.47\textwidth]{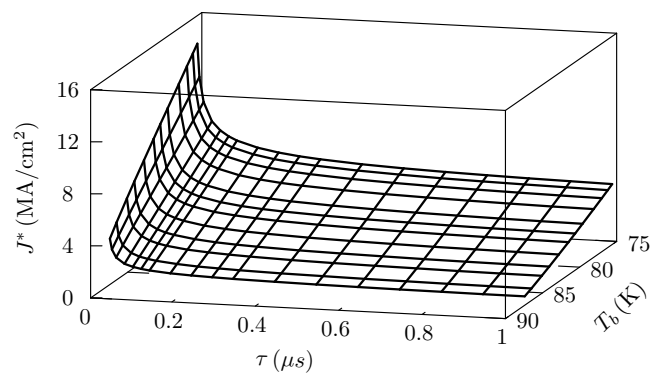}
\caption{\label{fig:TbJtau} $J^{*}$ as a function of bath temperature $T_{b}$
and the characteristic time of the film, $\tau$. }

\end{figure}

Intuitively, the plots for \emph{mA} and \emph{m50A} in Fig.\,\ref{fig:jstar}
differ because, the narrower the strip of film, the greater the proportion
of it that is effectively cooled via its edges as well as via its
bottom surface. It is therefore also to be expected that this edge
effect will only be significant for film strips narrower than a few
diffusion lengths. That this is so is indeed suggested by Fig.~\ref{fig:TbJtau},
which shows the surface $J^{*}(T_{b},\tau)$ calculated using $n$-isotherms
for sample \emph{mA}. For this superconductor and film thickness,
only films with characteristic times below about 0.05\,$\mu$s seem
likely to be almost free from thermal instability. More generally,
it appears that other mechanisms that limit superconductivity, such
as the Larkin-Ovchinnikov electrodynamic instability, should be studied
using films with very low $\tau$ (and/or very short intermittent
current pulses, as mentioned above) if the effect being studied is
not to be overwhelmed by the effect of thermal instability. 

\section{CONCLUDING REMARKS}

This paper has developed a very simple analytical model in which strips
of superconducting film supported by a substrate in a bath thermostatted
at temperature $T_{b}$ are represented as rectangular blocks embedded
in the surface of a semi-infinite medium of the same homogeneous material
as the block. Assuming a uniform current density in the block, and
given experimental current-voltage characteristics that allow parameterization
of the model, the space-averaged thermal dynamics of the film prove
to depend on its geometry through a single parameter, its characteristic
time $\tau$. The quantitative predictions of the theory as regards
the supercritical current density $J^{*}(T_{b})$ agree quite satisfactorily
with experimental observations (Fig.~\ref{fig:jstar}).

Though in this paper the model has been parameterized for and tested
on YBCO films 0.12$\,\mu\textrm{m}$ thick, its derivation involves
nothing that prevents its application to other high-T$_{c}$ or low-T$_{c}$
films, so long as they do not require refrigeration with superfluid
helium, the effective thermal conductivity of which is much greater
than that of any ceramic substrate\cite{Wilks87} even for boiling
films.\cite{Maza89} Whether or not the good performance shown in
Fig.\,\ref{fig:jstar} also generalizes to other films naturally
remains to be seen. However, the mutual similarity of the current-voltage
characteristics of all high-T$_{c}$ superconductors suggests that
they all become thermally unstable under high enough current densities,
and that, with appropriate parameterization, our model should be able
to predict this behaviour, thus explaining it in terms of thermal
instability. 

One of the key assumptions of the model is the homogeneity of the
superconducting film, which together with the homogeneity of current
density guarantees that the film and its behaviour are completely
described by its geometry and current-voltage characteristics. Since
YBCO films can have various degrees of inhomogeneity (mostly in relation
to their oxygen content),\cite{Qadri97} the successful application
of the model to \emph{mA} and \emph{m50A} suggests that the inhomogeneity
of these samples, if any, was sufficiently finely grained as to be
negligible. Further investigation of this issue probably requires
examination of numerous individual cases; certainly, it seems safe
to suppose that inhomogeneities can only  accelerate thermal runaway.

In this paper the intrinsic current-voltage characteristics of the
film, which are required for calculation of its heating rate [Eq.~(\ref{eq:elecpow})],
have been approximated by extrapolation from the low-energy region
of the experimental CVCs. Although the experimental CVCs suffer from
the very inaccuracies, the possible thermal origin of which is being
investigated, it is assumed that their low-energy regions coincide
sufficiently closely with the required intrinsic CVCs. That the functional
form used for extrapolation is not excessively critical is supported
in Table\,\ref{tab:J*m50a} by the agreement between $J_{n}^{*}$
and $J_{s}^{*}$ and the agreement of both with $J_{exp}^{*}$. Extrapolation
will not be necessary if ultra-fast current-voltage measurements with
nanosecond-scale measuring times become available, since such measurements
may be expected, for the reasons explained above, to be devoid of
thermal distortion. 

The most disconcerting feature of our model is no doubt its treating
the superconducting film and the substrate as a single continuum,
with a single diffusivity coefficient (the sole free parameter of
the model) and no acoustic mismatch between film and substrate. However,
the value of the diffusivity coefficient seems not to be excessively
critical (a single value worked well at all bath temperatures for
both the samples considered here); while the lack of acoustic mismatch
means that our results support the significance of thermal instability
effects even under the conditions that are least conducive to such
effects, i.e. with optimal thermal coupling between film and substrate.
Numerous authors have acknowledged the determinant role of thermal
effects when thermal impedance is high, but have implicitly or explicitly
denied that they are significant when thermal coupling is good.

To sum up, in previous studies,  finite element calculations\cite{Maza08}
have predicted the occurrence of a thermally driven transition to
the normal conductance state at zero applied magnetic field when a
homogeneous superconducting film is subjected to a controlled electrical
current exceeding a certain {}``supercritical'' value that coincides
with experimental observations.%
\footnote{When voltage instead of current is controlled the physics is quite
different. Heat input no longer increases monotonically with voltage
(which can lead to multivalued CVCs), and superconducting zones can
coexist with normal zones (e.g. due to hot spots). These phenomena
lie outside the scope of this paper.}

The present work supports those findings analytically. Experimental
research on the breakdown of superconductivity due to other mechanisms
(Larkin-Ovchinnikov vortex instabilities, hot spots, phase-slip centres)
must accordingly be carried out under conditions that exclude the
possibility of the heat-driven transition, and vice versa.

\section{Acknowledgements}

We gratefully acknowledge support by the Spanish Ministry of Science
and Innovation through contract ERDF FIS2010-19807 and by the Xunta
de Galicia through ERDF 2010/XA043 and 10TMT206012PR. 

Our special appreciation to Ian Coleman from the University of Santiago de Compostela for his decisive role in the English restyling of this paper.

\bibliographystyle{unsrt}
\bibliography{FilmStabil-Article3-OTHER}

\end{document}